\documentstyle[prb,aps,epsfig,twocolumn,graphics]{revtex}

\begin{document}
\draft

\tighten

\wideabs{
\title{The dynamics of a hole in a CuO$_4$ plaquette: 
       electron energy-loss spectroscopy of Li$_2$CuO$_2$}

\author{S.~Atzkern, M.~Knupfer, M.~S.~Golden, and J.~Fink} 
\address{Institut f\"ur Festk\"orper- und Werkstofforschung Dresden, 
         P.O.Box 270016, D-01171 Dresden, Germany} 

\author{C.~Waidacher, J.~Richter, and K.~W.~Becker}  
\address {Institut f\"ur Theoretische Physik, 
          Technische Universit\"at Dresden, D-01062 Dresden, Germany}

\author{N.~Motoyama, H.~Eisaki, and S.~Uchida} 
\address{Department of Superconductivity, The University of Tokyo, Bunkyo-ku, Tokyo 113, Japan} 

\date{\today}  
\maketitle

\begin{abstract}

We have measured the energy and momentum dependent loss function of
Li$_2$CuO$_2$ single crystals by means of electron energy-loss 
spectroscopy in transmission. Using the same values for the model parameters, the low-energy features of the spectrum as well as published Cu~2$p_{3/2}$ x-ray photoemission data of Li$_2$CuO$_2$ are well described by a cluster model that consists of a single CuO$_4$ plaquette only. This demonstrates that charge excitations in Li$_2$CuO$_2$ are strongly localized.
 \end{abstract}

}
\section{Introduction}

At present a large variety of cuprate compounds with different element
compositions and various crystal structures are known. The common 
structural unit of nearly all these materials is the square planar CuO$_4$ plaquette 
with the oxygen atoms at the corners and the copper atom in the center.
The  CuO$_4$ plaquettes can share oxygen atoms with their nearest neighbor 
plaquettes, thus building chains or planes. These Cu-O networks are separated generally
by counter ions. Therefore, the electronic properties of the cuprates
are low-dimensional. 

For a given doping level, the arrangement of the plaquettes in the crystal 
is decisive in determining the mobility 
and correlation of the charge carriers. Therefore, it influences the particular 
physical properties of the cuprate compound under consideration. For instance, in many two-dimensional (2D) 
systems high-T$_c$ superconductivity can be observed and among the quasi 
one-dimensional (1D) cuprates there exist excellent candidates for studying 
typically one-dimensional properties. Examples are the spin-Peierls transition which occurs in edge sharing chains 
(CuGeO$_3$)\cite{Hase93} or spin-charge separation in corner sharing chains
(Sr$_2$CuO$_3$).\cite{lieb68,fuji99,neud98} 
Furthermore, in the spin ladder compounds the ordering of the CuO$_4$ plaquettes lies
between 1D and 2D.  Their magnetic properties change discontinuously with the 
transition from 1D to 2D.\cite{dago96} One example of such a system is 
Sr$_{0.4}$Ca$_{13.6}$Cu$_{24}$O$_{41}$ which consists of 2-leg ladders as well as
chains of edge sharing plaquettes in alternating layers and which shows superconductivity 
under high pressure (3 GPa).\cite{ueha96}

In the hole picture (in the undoped case) every CuO$_4$ plaquette is occupied
by a single hole in otherwise empty Cu~3$d$ and O~2$p$ orbitals. Thus, in the case of interconnected CuO$_4$ plaquettes the 
theoretical analysis of the electronic structure 
has to deal with many-particle systems, and the interpretation of corresponding 
experimental data is rather complex.  Furthermore, it is usually not possible
to determine all theoretical model parameters for the Cu-O network from
experiments. For these reasons it is helpful to study in detail the
electronic properties of the structural unit, a single plaquette, itself to establish a basis for all Cu-O networks. 

On the other hand, in real materials plaquettes are never truly isolated. Therefore, we 
have to choose a system in which the electronic interaction between the holes 
on the plaquettes and charge carriers in their vicinity can be estimated to be 
small. One candidate for such a system is Li$_2$CuO$_2$. In view of the
structure of Li$_2$CuO$_2$ (see Fig.~\ref{structure}), one might assign this 
compound to the 1D systems due to the parallel aligned chains of edge-sharing 
plaquettes. However, since the Cu-O-Cu-bond angle along the chains is almost 
90$^\circ$, the hopping from one plaquette to its nearest neighbor  
is strongly suppressed. Thus, although the magnetic interactions in Li$_2$CuO$_2$
are rather complex\cite{Weht98,Boehm98,Yushankhai99,Mizuno99}, with respect to 
the electronic properties the plaquettes in this compound can be considered as 
approximately isolated. 

Recently, the optical properties and the electronic structure 
of Li$_2$CuO$_2$ have been studied carrying out reflectivity\cite{mizu98}, x-ray photoemission\cite{Boeske98} and x-ray absorption\cite{neud99}  measurements.
In this paper, we present transmission electron energy-loss spectroscopy (EELS) studies 
of Li$_2$CuO$_2$. The resulting spectra allow us to study the 
multipole transitions between the ground state and the excited states of an
approximately isolated single plaquette. Our theoretical analysis shows that
the EELS results, as well as the Cu 2$p_{3/2}$ x-ray photoemission spectrum
published by B\"oske et al.\cite{Boeske98}, can be described in a 
multi-band Hubbard model using the same parameter values for the 
charge-transfer energy and the hopping amplitudes.

\section{Experimental}

\begin{figure}
\begin{center}
\leavevmode \epsfig{file=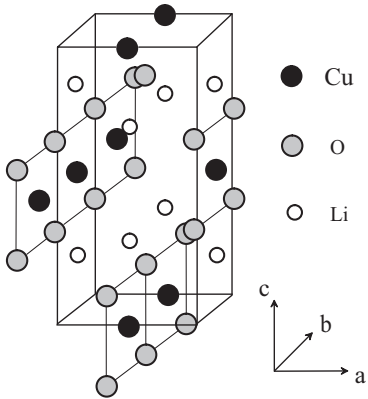,width=4.8cm,angle=0}
\end{center}
\caption{
Crystal structure of Li$_2$CuO$_2$. The Cu atoms (black filled circles), and 
their nearest neighbor O atoms (grey filled circles) build chains of edge
sharing CuO$_4$ plaquettes along {\bf b}. The Li atoms (open circles) are located 
right below and above the Cu atoms in {\bf c} direction.
\protect\label{structure}}
\end{figure}

Single crystals of Li$_2$CuO$_2$ were grown using a travelling 
solvent-zone technique.\cite{mizu98} The crystal structure is shown in 
Fig.~\ref{structure}. Li$_2$CuO$_2$ has a body centered orthorhombic unit cell 
(space-group Immm) with the lattice constants a~=~3.661~\AA , b~=~2.863~\AA , 
and c~=~9.393~\AA . Edge-sharing CuO$_4$ plaquettes lie in the ({\bf a},{\bf b}) plane 
and the chains are aligned along the
{\bf b} direction. \cite{Hoppe70,Sapina90} 
The Cu-O distance within the plaquette is 1.958~\AA\ while the O-O distances 
in {\bf b}- and {\bf c}-direction are 2.863~\AA\ and 2.671~\AA , respectively. 
This configuration implies a Cu-O-Cu-bond angle of about 94$^\circ$. The distances 
between a Cu in one chain and O atoms of neighboring chains are larger than 3~\AA .
This suggests that the electronic inter-chain coupling is small.

Electron energy-loss spectroscopy (EELS) in transmission was performed on free 
standing films of about 1000~\AA\ thickness which were cut from the crystals
using an ultramicrotome with a diamond knife. All measurements were carried out 
at room temperature and with a momentum transfer parallel to the {\bf b} and {\bf c} axis. 
The primary beam energy was 170~keV. The energy resolution 
$\Delta$E and the momentum transfer resolution $\Delta q$ were chosen to be 110~meV 
and 0.05~\AA $^{-1}$ for q~$\leq$~0.4~\AA $^{-1}$, and 160~meV and 0.06~\AA $^{-1}$   
for q~$>$~0.4~\AA $^{-1}$. 

EELS provides us with the energy and momentum dependent loss function 
Im(-1/$\varepsilon(\omega ,{\bf q}$)) which is directly proportional to the 
imaginary part of the dynamical density-density correlation function
\begin{equation}
\chi_{\rho}(\omega,{\bf q}) = \frac{1}{i} \int_0^{\infty} 
dt~e^{-i\omega t}\langle\Psi|
[ \rho_{-{\bf q}}(0),\rho_{\bf q}(t)]|\Psi\rangle~\mbox{.}\label{correlation}
\end{equation}
$|\Psi\rangle$ describes the ground state of the Hamiltonian, and $\rho_{{\bf q}}$ is the 
Fourier transformed hole density (see below). Equation~(\ref{correlation})
implies that for the limit {\bf q}~$\rightarrow 0$ the selection rules for
transitions are the same as in optics, i.e. dipole transitions are allowed. For finite {\bf q} 
non-dipole transitions additionally contribute to the loss function. 

\section{Results and Discussion}

\begin{figure}
\begin{center}
\leavevmode \epsfig{file=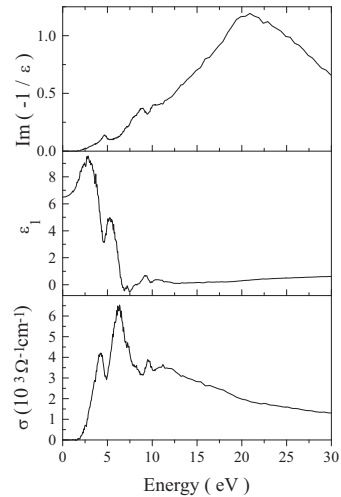,width=5.5cm,angle=0}
\end{center}
\caption{
Loss function, Im(-1/$\varepsilon$) (upper panel), real part of the dielectric function, $\varepsilon_1$ (middle panel), and optical conductivity, $\sigma = \omega\varepsilon_0\varepsilon_2$ (lower panel) of  Li$_2$CuO$_2$ for a small momentum transfer of 0.08~\AA $^{-1}$ along the {\bf b} direction. 
\protect\label{kka}}
\end{figure}

In Fig.~\ref{kka} we show the loss function as well as the real part of the dielectric function, $\varepsilon_1$, and the optical conductivity, $\sigma = \omega\varepsilon_0\varepsilon_2$, of Li$_2$CuO$_2$ for a momentum transfer of {\bf q} = 0.08~\AA$^{-1}$. The latter two have been derived by a Kramers-Kronig analysis. The loss function is dominated by a broad feature around 21~eV which represents the volume plasmon, a collective excitation of all valence electrons. At lower energies several further plasmon excitations can be observed which are related to transitions between occupied and unoccupied electronic states with mainly Cu 3$d$, Cu 4$s$, and O 2$p$ character . 
These transitions give rise to the maxima in the optical conductivity which is shown in the lower panel of Fig.~\ref{kka}. At small momentum transfers $\sigma$ as derived from EELS can be directly compared to results from optical measurements. The optical conductivity of Li$_2$CuO$_2$ shown in Fig.~\ref{kka} is in good agreement with recent optical studies.\cite{mizu98} Note that the corresponding maxima in the loss function are observed at higher energies when compared to the optical conductivity. This is a direct consequence of the collective nature of the plasmon excitations as observed in the loss function.\cite{fink}

\begin{figure}[tbh]
\begin{center}
\leavevmode \epsfig{file=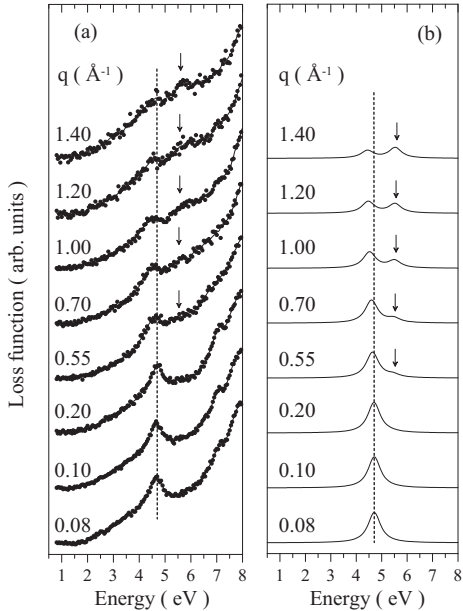,width=6.5cm,angle=0}
\end{center}
\caption{
Loss function of Li$_2$CuO$_2$ (left panel) measured with different momentum 
transfers {\bf q} parallel to the chain direction. The right panel shows the calculated 
dynamical density-density correlation function for the same {\bf q} values. The dashed lines and the arrows illustrate the small dispersion of the main peak and the increase of spectral weight with increasing {\bf q} around 5.5~eV, respectively.
\protect\label{eels}}
\end{figure}

Fig.~\ref{eels}a focusses on the loss function in a smaller energy range as a function of the momentum transfer. 
For small momentum transfer the energy range up to 6~eV is dominated by one 
distinct peak at 4.7~eV. 
With increasing {\bf q} the intensity of this peak 
decreases while the intensity of a second feature around 5.7~eV, which becomes 
visible at q $\geq$ 0.7~\AA$^{-1}$, increases. This behavior of the two features can 
be explained by the different multipole character of the corresponding 
transitions (see below). The steep rise of the background above 6~eV is due to the higher lying excitations (see Fig.~\ref{kka}).

Since the Cu-O-Cu-bond angle is close to 90$^\circ$, the nearest-neighbor (NN) 
Cu-O-Cu hopping amplitude is small, and the delocalization properties of 
the system are dominated by a weak next-nearest neighbor (NNN) Cu-O-O-Cu 
hopping.\cite{Weht98} In the loss function the NN as well as the NNN hopping 
might be visible in the tail of non-vanishing spectral weight between the 
spectral onset at $\sim$ 1.5 eV and the main peak at 4.5 eV as has been discussed previously.\cite{mizu98}  
However, in a first approximation the CuO$_4$ plaquettes can be regarded as 
isolated and the charge excitations in the Cu-O network of Li$_{2}$CuO$_{2}$ 
as localized. This is supported by the small dispersion of the features in the EELS spectra  which is in contrast to the strong  positive dispersion of about 0.5 eV of the lowest lying excitations e.g. in the 1D system Sr$_2$CuO$_3$\cite{neud98} and in the 2D system Sr$_2$CuO$_2$Cl$_2$.\cite{wang96,neud97} In addition, the fact that the spectra for {\bf q} parallel to the {\bf c} direction (not shown) are essentially identical to those for 
{\bf q} parallel to the {\bf b} direction also demonstrates the zero-dimensional character of the excitations. 

\begin{figure}[tbh]
\begin{center}
\leavevmode \epsfig{file=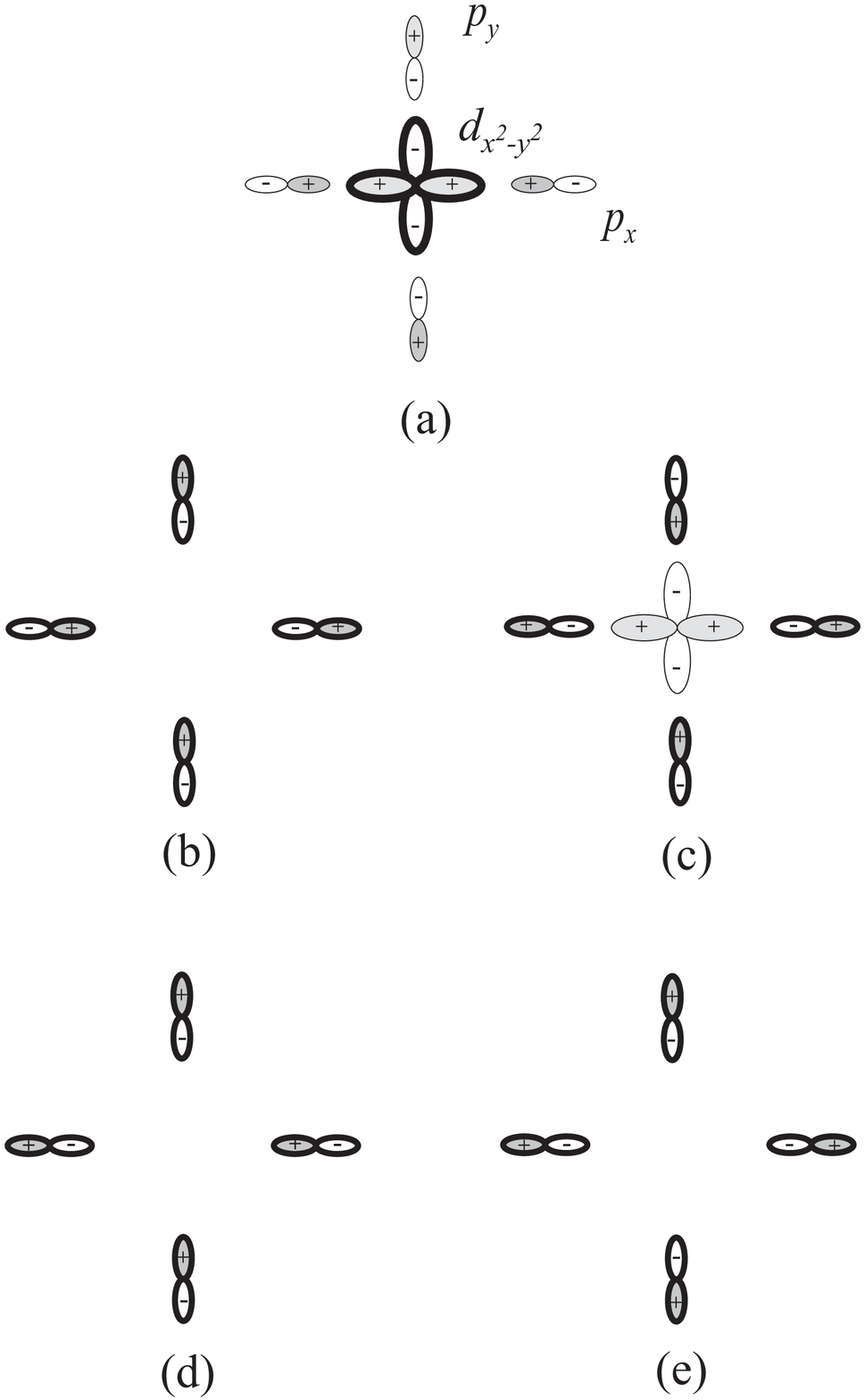,width=6cm,angle=0}
\end{center}
\caption{
Symmetry properties of the ground state $\left(\text{a}\right)$, 
and the final states $\left(\text{b}\right)$-$\left(\text{e}\right)$ of Hamiltonian(\protect\ref{Hamiltonian}).
The relevant combinations of the hole occupied orbitals (Cu $3d_{x^2-y^2}$ and O 2$p_{x,y}$) in the plaquette are shown.
Thicker lines correspond to larger hole densities. The states
$\left(\text{b}\right)$, $\left(\text{d}\right)$ and $\left(\text{e}\right)$ have zero Cu weight. 
For momentum transfer parallel to the chain 
direction non-vanishing contributions come only from the final states 
$\left(\text{b}\right)$ and $\left(\text{c}\right)$.
\protect\label{eigenstates}}
\end{figure}

Consequently, our model system for the calculation of the loss functions
consists of a single CuO$_{4}$ plaquette only. This is a cluster with one Cu $3d_{x^2-y^2}$ 
orbital and four O $2p_{x(y)}$ orbitals (with site index $j=1,\ldots 4$), 
which is occupied by a single hole. The restriction to solely these orbitals is fully consistent with x-ray absorption measurements which indicate that the highest occupied
states in Li$_{2}$CuO$_{2}$ have only small hole density in orbitals perpendicular to the CuO$_4$ plaquette.\cite{neud99}
In this case, $\rho_{{\bf q}}$
has the form
\begin{equation}
\rho_{{\bf q}} = n^d +\sum_{j} n_{j}^p e^{i{\bf q}{\bf r}_j}
~\mbox{,}\label{density} 
\end{equation}
and the Hamiltonian reads
\begin{eqnarray}
H &=&\Delta \sum_{j}n_{j}^p
+t_{pd}\sum_{j}\phi _{j}\left(p_{j}^{\dagger }d
+d^{\dagger }p_{j}\right)\nonumber\\
&&+ t_{pp}\sum_{\left\langle j,j^{\prime }\right\rangle}
\phi _{jj^{\prime }}\,p_{j^{\prime}}^{\dagger }p_{j}
~\mbox{.}\label{Hamiltonian}
\end{eqnarray}
$d^{\dagger }$ $\left( p_{j}^{\dagger }\right)$ creates a hole in
the Cu $3d$ orbital ($j$-th O $2p$ orbital), with occupation number 
operators $n^d$ ($n_{j}^p$). $\Delta $ is the charge-transfer energy. 
There are two hopping parameters: $t_{pd}$ for the Cu-O hopping, 
and $t_{pp}$ for the O-O hopping. The signs for the hopping are
defined as follows: $\phi _{1}=\phi _{2}=-1$, $\phi _{3}=\phi _{4}=1$, 
and $\phi _{jj^{\prime }}=-\phi _{j}\phi _{j^{\prime }}$. $\langle
jj^\prime\rangle$ denotes the summation over nearest neighbor pairs. 
Hamiltonian (\ref{Hamiltonian}) is the simplest model which captures
the essential features of the system. It can easily be solved
exactly. 
The inclusion of anisotropic hopping will be discussed below. 

Figure~\ref{eigenstates} shows the symmetry properties of the ground state
$\left(\text{a}\right)$, and the four excited states 
$\left(\text{b}\right)$-$\left(\text{e}\right)$ of model (\ref{Hamiltonian}). 
The ground state is a $\sigma$-bonding combination of the Cu 3$d_{x^2-y^2}$ 
orbital and the four O 2$p$ orbitals. Transitions into states 
$\left(\text{c}\right)$ and $\left(\text{e}\right)$ are optically forbidden.
Thus, they only contribute to the loss function for finite momentum 
transfer. Moreover, if ${\bf q}$ is parallel to the chain direction, non-vanishing 
contributions arise only from transitions into the final states 
$\left(\text{b}\right)$ and $\left(\text{c}\right)$. State
$\left(\text{b}\right)$ has pure O 2$p$ character, whereas state 
$\left(\text{c}\right)$ is the anti-bonding counterpart to 
$\left(\text{a}\right)$. For $\mathbf{q}$ = 0 dipole transitions 
into state $\left(\text{b}\right)$ lead to spectral weight at energy 
$\omega _{1}$. A second excitation at energy $\omega _{2}$
due to transitions into state $\left(\text{c}\right)$ has non-vanishing weight 
for $\mathbf{q}\neq$ 0: 
\begin{eqnarray}
\omega _{1} &=&\frac{\Delta +2t_{pp}+\omega _{2}}{2}
~\mbox{,}\label{excitation1} \\ 
\omega _{2} &=&\sqrt{\left( \Delta -2t_{pp}\right) ^{2}+16t_{pd}^{2}}
~\mbox{.}\label{excitation2}
\end{eqnarray}

These transitions, with energies $\omega_{1}$ and $\omega_{2}$, appear 
as plasmon excitations in the EELS spectra at somewhat higher energies $\omega^{\text{RPA}}_{1}$ and 
$\omega^{\text{RPA}}_{2}$. 
This effect is a result of the long-range Coulomb interaction (see also the discussion above).
In the theoretical model this issue can be satisfactorily treated within the random-phase approximation (RPA).\cite{Pines63} 

From the experimental spectrum we obtain for small momentum transfer
$\omega^{\text{RPA}}_{1}=4.7$~eV (see Fig.~\ref{eels}a). For larger 
$\mathbf{q}$ the experimental spectra show firstly a small negative dispersion of the main peak (see the deviation from the vertical dashed line in Fig.~\ref{eels}a) amounting to about 0.3~eV in the range between 0.1  and 1.4~\AA$^{-1}$.
This can be assigned to the decrease of the long-range Coulomb potential and the decreasing oscillator strength of the transition with increasing {\bf q}.
Secondly, the spectral weight between $5.5$ and $5.8$~eV (in Fig.~\ref{eels}a marked by arrows) is enhanced 
at higher q, resulting from the plasmon excitation related to the dipole forbidden transition at $\omega_{2}$.
Since the excitation into the final state 
$\left(\text{c}\right)$ corresponds to an octupole transition, the intensity 
of the corresponding feature in the EELS spectra is low at small {\bf q}, and becomes visible 
only at rather high momentum transfers. 

The two excitation energies $\omega_{1}$ and $\omega_{2}$ in equation (\ref{excitation1}) and (\ref{excitation2}) are described by the charge transfer energy $\Delta$ and the two hopping parameters $t_{pd}$ and $t_{pp}$. Thus the energy positions of the two peaks observed in the EELS spectra do not yet determine the parameter set in a unique way. In order to remove the last degree of freedom in our
model we use experimental evidence from Cu $2p_{3/2}$ core-level x-ray
photoemission spectroscopy (XPS). Thereby we take advantage of the fact that three additional measured parameters in the XPS spectra, namely the energy positions as well as the ratio of the spectral weights of the main and satellite line are obtained, whereas only two variables are added to our model, namely the Coulomb repulsion $U_{dc}$ and the exchange parameter $I_{dc}$. As explained in Ref.~\onlinecite{Waidacher99} 
for other compounds, we use model (\ref{Hamiltonian}) to calculate the core 
level XPS of Li$_{2}$CuO$_{2}$. Good agreement with the experimental 
spectra\cite{Boeske98} is obtained if $t_{pd}^{2}/\left(\Delta -2t_{pp}\right)$ 
lies between $0.7$ and $0.8$~eV.

It turns out that both the EELS and the XPS results can be described 
using one single parameter set for Hamiltonian (\ref{Hamiltonian}):
$\Delta=2.7~\mbox{eV}$, $t_{pd}= 1.25~\mbox{eV}$, and $t_{pp}=0.32~\mbox{eV}$\cite{udc} 
which thus represents the most reliable parameter values for a CuO$_{4}$ plaquette in Li$_{2}$CuO$_{2}$. Our values for the charge-transfer energy $\Delta$ and the O-O hopping $t_{pp}$ lie 
between those obtained from band-structure calculations\cite{Weht98} 
($\Delta=2.5~\mbox{eV}$, $t_{pp}=0.25~\mbox{eV}$), and from a fit to optical 
measurements\cite{mizu98} ($\Delta=3.2~\mbox{eV}$, $t_{pp}=0.56~\mbox{eV}$), while 
the value for $t_{pd}$ is about $10\%$ larger than in Refs.~\protect\onlinecite{Weht98}
and~\protect\onlinecite{mizu98}.

\begin{figure}[tbh]
\begin{center}
\leavevmode \epsfig{file=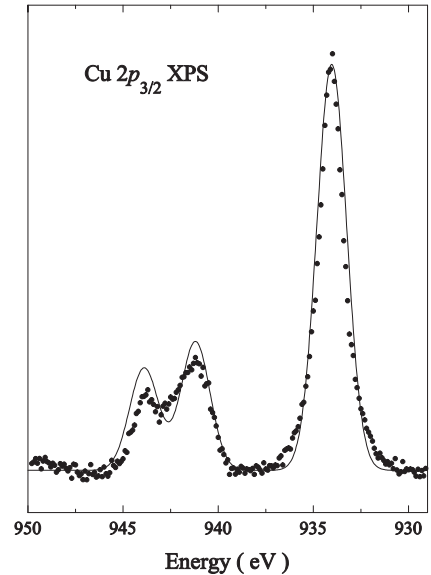,width=5.5cm,angle=0}
\end{center}
\caption{
Cu $2p_{3/2}$ photoemission spectrum of  Li$_2$CuO$_2$, taken from 
Ref.~\protect\onlinecite{Boeske98}. The solid line shows the theoretical 
results. Details of the calculation are described in 
Ref.~\protect\onlinecite{Waidacher99}.
\protect\label{xps}}
\end{figure}

In Figs.~\ref{eels}b and \ref{xps} the theoretical results are compared 
to the experimental  EELS and XPS spectra. The theoretical 
spectra have been artificially broadened with a full width at half maximum 
of $0.6~\mbox{eV}$ for the EELS (Fig.~\ref{eels}b) and $1.8~\mbox{eV}$ for 
the XPS (Fig.~\ref{xps}). In agreement with the experimental result 
(Fig.~\ref{eels}a) the calculated EELS data for small momentum transfer consist of a
single peak at $\omega^{\text{RPA}}_{1}=4.7$ eV. The energy $\omega_{1}$ = 4.4~eV, calculated with the parameter set received from our data, agrees well with the peak position in the optical conductivity as obtained performing a Kramers-Kronig analysis (see Fig.~\ref{kka}) and as derived from optical studies.\cite{mizu98}  For increasing momentum
transfer, the peak in the calculated loss function shows a small negative dispersion which is also in full agreement with the experiment. Furthermore, a second
feature appears around $\omega^{\text{RPA}}_{2}=5.5$ eV, the symmetry
properties of the associated final states have been discussed above. 
As regards the XPS results (Fig.~\ref{xps}), we also find a good agreement between theory 
and experiment. The structure of
the experimental spectrum, which consists of a broad satellite (at higher
binding energies) and a narrow main line, is correctly reproduced. In addition, 
the theoretical value for the ratio of the spectral weights of the satellite and the main line coincides with the experimental intensity ratio (0.56).\cite{Boeske98}

We conclude with a short discussion of anisotropic hopping. In Li$_{2}$CuO$_{2}$, 
the O-O distance in chain direction ({\bf b} direction) is about $7\%$ larger 
than in {\bf a} direction.\cite{Sapina90} Thus, one 
expects an anisotropy in the O-O hopping, with a larger hopping parameter 
$t_{pp}^{\perp }$ perpendicular to the chain direction (see also Ref.~\onlinecite{mizu98}). However, if ${\bf q}$ is 
parallel to the chain direction, a larger $t_{pp}^{\perp }$ in the present model
can be absorbed into a renormalized charge-transfer energy $\Delta$ without 
affecting the spectra. As the anisotropy of the edge-sharing CuO$_4$ plaquettes results in a finite
Cu-O-Cu hopping to the next CuO$_4$ unit, an increase of this anisotropy causes 
a step towards a one-dimensional electronic system. This is actually realized 
in the related spin-Peierls compound CuGeO$_3$ where the anisotropy is about 16\%. Thus, CuGeO$_3$ might be an ideal candidate for the investigation of the electronic structure of a system at the border to one dimension.

\section{Summary}

In summary, we have carried out EELS measurements of Li$_2$CuO$_2$ single
crystals. Using the same model with the same parameter values, calculations of the energy and momentum dependent loss function of an isolated  CuO$_4$ plaquette as well as of the Cu~2$p_{3/2}$ core-level XPS of Li$_2$CuO$_2$ agree well with the experimentally observed low-energy features of EELS as well as with published XPS data, respectively.
Thus, a single parameter set for the
charge-transfer energy ($\Delta$~=~2.7~eV), the copper-oxygen
(t$_{pd}$~=~1.25~eV) and oxygen-oxygen (t$_{pp}$~=~0.32~eV) hopping 
amplitudes is sufficient to describe two independent experiments probing the electronic 
structure of Li$_{2}$CuO$_{2}$ in
completely different energy ranges. These results confirm the strongly
localized character of charge excitations in Li$_{2}$CuO$_{2}$.

This work was supported by DFG through the research program of the SFB 463, Dresden. 



\newpage






\begin{references}
\bibitem{udc}
        The influence of the core hole in the XPS has been modelled using a Coulomb repulsion
$U_{dc}=8.35~\mbox{eV}$ and an exchange parameter $I_{dc}=-1.5~\mbox{eV}$, 
consistent with other cuprates (see Ref.~\protect\onlinecite{Waidacher99}).
\end{references}
\end{document}